%% file: main.tex
\documentclass[sigplan,10pt]{acmart}
\usepackage{tikz}
\usetikzlibrary{positioning}
\usepackage{algorithm}
\usepackage{algpseudocode}
\usepackage{placeins}

\acmConference{}{}{} 

\acmYear{February 2025}
\acmISBN{} 
\acmDOI{} 

\setcopyright{none}

\usepackage{booktabs}   
\usepackage{subcaption} 
\include{mac-includes}

\usepackage[normalem]{ulem}

\usepackage{bm}

\usepackage{graphicx}
\usepackage{textcomp}
\def\BibTeX{{\rm B\kern-.05em{\sc i\kern-.025em b}\kern-.08em
    T\kern-.1667em\lower.7ex\hbox{E}\kern-.125emX}}

\usepackage{fontawesome}
\usepackage{pifont}
\usepackage{textcomp}

\usepackage{soul}

\usepackage{dblfloatfix}

\usepackage{mathrsfs}

\usepackage{algorithm,algorithmicx,algpseudocode}

\usepackage{tikz}
\usetikzlibrary{tikzmark}

\usepackage{xpatch}
\expandafter\xpatchcmd
\csname pgfk@/tikz/every picture/.@cmd\endcsname
{\thepage}{\arabic{page}}{}{}

\tikzstyle{comment} = [draw, fill=blue!70, text=white, text width=3cm, minimum height=1cm, rounded corners, align=left, font=\scriptsize]
\tikzstyle{background_alg} = [draw, fill=blue!20, opacity=0.4, inner sep=4pt, rounded corners=2pt]

\definecolor{vlgray}{rgb}{0.77 0.77 0.77}
\definecolor{ablack}{rgb}{0.2 0.2 0.2}
\definecolor{grayblack}{rgb}{0.4 0.4 0.4}

\usepackage[customcolors]{hf-tikz}
\hfsetbordercolor{white}
\hfsetfillcolor{vlgray}

\newcounter{highlight}

\newcounter{Ahighlight}

\newif\ifrev
\revfalse

\sethlcolor{white}

\begin{document}

\title[]{Performance-Driven Optimization of Parallel Breadth-First Search}



\author{Marati Bhaskar}
\orcid{nnnn-nnnn-nnnn-nnnn}             
\affiliation{
  \department{Department of Computer Science and Engineering}              
  \institution{IIT Tirupati}            
  \country{India}                    
}
\email{cs24d001@iittp.ac.in}         

\author{Raghavendra Kanakagiri}
\orcid{nnnn-nnnn-nnnn-nnnn}             
\affiliation{
  \department{Department of Computer Science and Engineering}              
  \institution{IIT Tirupati}            
  \country{India}                    
}
\email{raghavendra@iittp.ac.in}          

\begin{abstract}
Breadth-first search (BFS) is a fundamental graph algorithm that presents significant challenges for parallel implementation due to irregular memory access patterns, load imbalance and synchronization overhead. 
In this paper, we introduce a set of optimization strategies for parallel BFS on multicore systems, including hybrid traversal, bitmap-based visited set, and a novel non-atomic distance update mechanism.
We evaluate these optimizations across two different architectures - a 24-core Intel Xeon platform and a 128-core AMD EPYC system - using a diverse set of synthetic and real-world graphs. Our results demonstrate that the effectiveness of optimizations varies significantly based on graph characteristics and hardware architecture. For small-diameter graphs, our hybrid BFS implementation achieves speedups of $3-8\times$ on the Intel platform and $3-10\times$ on the AMD system compared to a conventional parallel BFS implementation. However, the performance of large-diameter graphs is more nuanced, with some of the optimizations showing varied performance across platforms including performance degradation in some cases. 
\end{abstract}



\maketitle

\input{introduction}
\input{optimizations}

\input{evaluation}
\input{conclusion}

\bibliography{bibfile}



\end{document}

%% file: mac-includes.tex
\usepackage{etex}

\usepackage[font={sf,scriptsize}]{caption}

\usepackage{makecell,cellspace}
\usepackage{graphicx}
\graphicspath{{figs/},{plots/}}
\usepackage{booktabs}
\usepackage{epstopdf}
\usepackage{url}
\usepackage{placeins}

\usepackage{amsmath,amssymb,amsfonts,amsthm}
\usepackage{mathtools,mathrsfs}
\usepackage{multirow}
\usepackage{rotating}
\usepackage{pbox}
\usepackage[normalem]{ulem}
\usepackage{hanging}

\usepackage{listings}

\usepackage{etoolbox}

\makeatletter
\patchcmd{\ttlh@hang}{\parindent\z@}{\parindent\z@\leavevmode}{}{}
\patchcmd{\ttlh@hang}{\noindent}{}{}{}
\makeatother

\usepackage[utf8]{inputenc}
\newcommand{\cref}[1]{\ref{#1}}

\usepackage[10pt]{moresize}

\usepackage{enumitem}

\usepackage{xcolor}
\definecolor{darkgrey}{RGB}{140,140,140}
\definecolor{lightgrey}{RGB}{200,200,200}
\definecolor{lblue}{RGB}{0,76,153}
\definecolor{lightblue}{RGB}{173,216,230}
\definecolor{darkred}{RGB}{139,0,0}

\usepackage{scalerel,stackengine}
\stackMath
\newcommand\rwh[1]{%
\savestack{\tmpbox}{\stretchto{%
  \scaleto{%
      \scalerel*[\widthof{\ensuremath{#1}}]{\kern-.6pt\bigwedge\kern-.6pt}%
          {\rule[-\textheight/2]{1ex}{\textheight}}
            }{\textheight}%
}{0.5ex}}%
\stackon[1pt]{#1}{\tmpbox}%
}

%% file: introduction.tex
\section{Introduction}

Graphs are data structures that represent relationships between entities, with vertices denoting entities and edges capturing their connections. Graph algorithms analyze these relationships and are crucial in applications ranging from social networks to bioinformatics. Breadth-first search (BFS) is a fundamental graph algorithm that explores vertices level by level from a source vertex, serving as a building block for many other graph algorithms.

Despite its simple design, parallelizing BFS efficiently is challenging due to irregular memory access patterns, load imbalance and synchronization overheads. While several works have explored BFS optimizations across different architectures using various techniques~\cite{think_like_a_vertex, ligra, gunrock, enterprise, graphblast, pasgal,beamer:direction-optimizing}, achieving high performance remains difficult. In this paper, we investigate various optimization strategies for parallel BFS on multicore systems. We evaluate these optimizations across two different architectures and analyze their impact on BFS performance using diverse graph datasets.

In this paper, we make the following contributions:
\begin{itemize}
    \item We introduce a novel optimization strategy that uses non-atomic updates for distance values in BFS kernels. We quantify the opportunities for non-atomic updates and evaluate the impact on performance.
    \item We propose a simple yet effective heuristic for dynamically switching between top-down and bottom-up BFS phases based on the frontier size. 
    \item We present a bitmap-based approach for tracking visited vertices in BFS, which reduces memory overheads and improves cache locality.
    \item We evaluate our optimizations on two different platforms: SpeedCode~\cite{speedcode} and an in-house x86 server, and show that the effectiveness of optimizations is influenced by both graph characteristics and hardware architecture.
\end{itemize}

%% file: optimizations.tex
\section{Optimization Strategies}
\label{sec:optimizations}

In this section, we describe several optimizations to improve the performance of parallel BFS on multicore processors. We present these optimizations in the context of the parallel BFS kernel shown in Listing~\ref{lst:bfs_v1} (excluding the highlighted code block in blue). We refer to this kernel as the \texttt{BFS-Conventional} kernel, which uses atomic operations to ensure single-threaded updates to vertex distances.

Throughout our discussion of optimizations, the term `local frontier' refers to the vertices collected by each individual thread during an iteration, before they are merged into the global frontier. While we experimented with various parallel scheduling approaches, including dynamic scheduling with different chunk sizes, static scheduling consistently showed the best average performance on the Speedcode platform~\cite{speedcode} in our experiments.

~\\
\noindent \textbf{\texttt{BFS-NonAtomic:}} In the \texttt{BFS-Conventional} kernel, we use atomic operations to update vertex distances. These atomic operations are expensive and can lead to contention. In our bulk-synchronous BFS implementation, since threads update vertex distances synchronously level by level, a vertex's distance is updated with the same value even when multiple threads attempt to update it. Removing atomic updates introduces two issues: (1) Duplicate work: The same vertex can be added to the frontier multiple times by different threads, leading to redundant processing in the next iteration, (2) Data races: Since multiple threads update the same memory location without synchronization, this creates a data race according to the C++11 standard, which defines it as concurrent conflicting actions where at least one is non-atomic. (Section 1.10.21 in~\cite{cpp:standard}). While multiple threads updating the same memory location without synchronization is technically undefined behavior according to the C++ standard, our implementation exploits the fact that all concurrent updates write the same distance value due to the level-synchronized nature of BFS. In Section~\ref{sec:evaluation}, we quantify the performance impact of this optimization and caution that its usage should be limited to architectures that provide the necessary hardware guarantees and contexts where compiler optimizations won't affect the intended behavior. Also note that this optimization is specific to BFS's distance updates and cannot be employed if a node's data must be strictly updated only once.

\begin{figure*}
    \centering
    \includegraphics[scale=0.41]{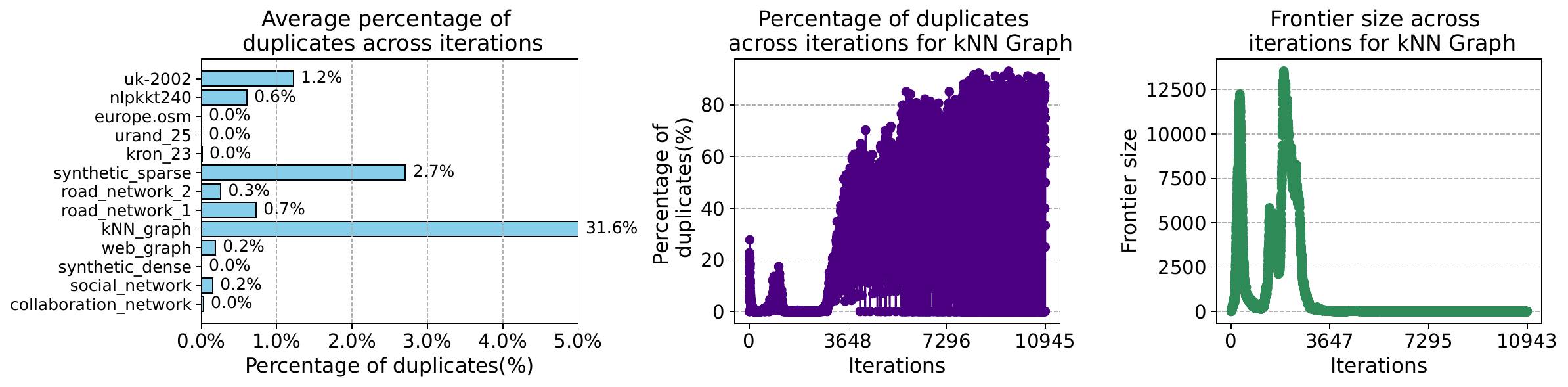}
    \vspace{-2mm}
    \caption{Analysis of duplicates and frontier size across iterations.}
    \label{fig:avg_duplicates}
\end{figure*}

In Figure~\ref{fig:avg_duplicates} (left), we show the average percentage of duplicates in the frontier across iterations without atomic updates. We observe that the percentage of duplicates is less than $1\%$ of the total frontier size across several datasets. The dataset \texttt{kNN\_graph} has the highest average duplicates at $31.6\%$. Figures~\ref{fig:avg_duplicates} (middle and right) show the percentage of duplicates and frontier sizes across iterations for the \texttt{kNN\_graph}, respectively. We observe that the percentage of duplicates increases in later iterations when the frontier size becomes small (less than $1\%$ of total vertices). Therefore, although the percentage of duplicates appears high, the actual number of duplicate vertices is small. Since many threads would be idle during these small-frontier iterations anyway, having them process duplicate vertices does not negatively impact performance.

\begin{lstlisting}[aboveskip=0em, belowskip=0em, float=!h,label=lst:bfs_v1,caption=Parallel BFS kernel showing both atomic and non-atomic distance updates.]
#pragma omp parallel
{
  while (cfsz != 0) {
    // Traverse all vertices in current frontier
    #pragma omp for nowait
    for (uint32_t i = 0; i < cfsz; ++i) {
      vidType src = curr_frontier[i];
      // Traverse all neighbors of frontier vertex
      for (uint32_t j = rptr[src]; j < rptr[src+1]; ++j) {
        vidType dst = col[j];
        // if neighbor is not visited
        if (dist[src] + 1 < dist[dst]) {
          // |\color{darkred}{BFS-Conventional: atomically update the neighbor's distance}|
          |\vspace{-0.8em}|
          |\vspace{0.8em} \tikzmarkin[set fill color=lightgrey, set border color=white, above offset=0.22, below offset=-0.1, right offset=2.45]{codebox}
          if (\_\_sync\_bool\_compare\_and\_swap(\&dist[dst], MAX\_DISTANCE, dist[src] + 1)) \{
          \tikzmarkend{codebox} |

          // |\color{darkred}{BFS-NonAtomic: Update the distance directly. See issue 2 as detailed in \texttt{BFS-NonAtomic}}|
          |\vspace{-0.8em}|
          |\vspace{0.3em} \tikzmarkin[set fill color=lightblue, set border color=white, above offset=0.22, below offset=-0.1, right offset=3.62]{codebox3}
          dist[dst] = dist[src] + 1;
          \tikzmarkend{codebox3} |

            // Add to local frontier
            lf[lfsz++] = dst;
          |\vspace{0.6em} \tikzmarkin[set fill color=lightgrey, set border color=white, above offset=0.22, below offset=-0.1, right offset=6.6]{codebox2}
          \}
          \tikzmarkend{codebox2} |
        }
      }
    }
    if (lfsz != 0) {
      // collect local frontiers
      fstart = __atomic_fetch_add(&nxt_fsz, lfsz, __ATOMIC_SEQ_CST);
      for (uint32_t i = 0; i < lfsz; ++i)
        next_frontier[fstart+i] = lf[i];
    }
    #pragma omp barrier
    
    #pragma omp single
    {
      swap(curr_frontier, next_frontier);
      cfsz = nxt_fsz; nxt_fsz = 0;
    }
    lfsz = 0;
  }
}
\end{lstlisting}

~\\
\noindent \textbf{\texttt{BFS-Hybrid:}} In~\cite{beamer:direction-optimizing}, the authors propose a hybrid BFS algorithm that dynamically switches between top-down and bottom-up phases based on frontier size. In general, the top-down phase is used when the frontier is small, whereas the bottom-up phase is employed when it is large. In the bottom-up phase, for each unvisited vertex, the algorithm checks if any of its neighbors are in the current frontier. If so, the unvisited vertex's distance is updated and it's added to the next frontier. This optimization reduces the number of edges traversed. The switching criteria in~\cite{beamer:direction-optimizing} is based on the next frontier edges to be visited ($m_f$), total unvisited edges ($m_u$), and next frontier size ($f_{i+1}$). The algorithm switches from top-down to bottom-up when $m_f > m_u / \alpha$, and switches back when $f_{i+1} < N / \beta$ and $f_{i+1} < f_i$. $\alpha$ and $\beta$ are tuning parameters.
We refer to this as $\texttt{BFS-Hybrid}^{\dagger}$. We simplify these criteria to use only the next frontier size, eliminating the need to compute and synchronize $m_f$ and $m_u$ across threads. Our version, which we call $\texttt{BFS-Hybrid}$, switches phases when the frontier size crosses $5\%$ of the total vertices and is detailed in Listing~\ref{lst:bfs_v3}.
\begin{lstlisting}[aboveskip=0em, belowskip=0em, float=!h,label=lst:bfs_v3,caption=Hybrid parallel BFS kernel with dynamic switching between top-down and bottom-up phases]
is_top_down = true;
#pragma omp parallel
{
  while (cfsz != 0) {
    if(is_top_down) {
      // Same as |\text{line 5-32} of Listing: \ref{lst:bfs_v1}$  \texttt{ BFS-NonAtomic}$|
    }
    else{
      // Frontiers are collected in a bitmap
      // Reset nxt_front_bm
      #pragma omp for nowait reduction(+: nxt_fsz)
      for (vidType v = 0; v < N; v++) { 
        if (dist[v] == MAX_DISTANCE) { 
          // Check if its neighbors are in the current frontier
          for (uint32_t j = rptr[v]; j < rptr[v+1]; ++j) {
            vidType neighbor = col[j];
            if (curr_front_bm[neighbor]) {
              dist[v] = depth + 1;
              nxt_fsz++;
              nxt_front_bm[v] = true;
              break;
            }
          }
        }
      }
    }
    #pragma omp barrier
    #pragma omp single
    { 
      // update curr_frontier, curr_front_bm, cfsz and nxt_fsz
      is_top_down = (cfsz < 0.05 * N)? true : false;
      depth++;
    }
    #pragma omp barrier
    if(move_from_TD_to_BU) {
      // convert  frontier array to curr_front_bm (bitmap)
    }
    else if(move_from_BU_to_TD) {
      // convert curr_front_bm (bitmap) to curr_frontier (array)
    }
    lfsz = 0;
    #pragma omp barrier
  }
}
\end{lstlisting}

~\\
\noindent \textbf{\texttt{BFS-VisitedBitmap:}} The \texttt{BFS-Hybrid} relies on the distance array to check visited vertices in both phases. These random accesses to the distance array can lead to cache misses. Moreover, loading a 4- or 8-byte integer from memory just to check if a vertex is visited is inefficient. We propose using a bitmap to track visited vertices instead, an optimization we call \texttt{BFS-VisitedBitmap}. Listing~\ref{lst:bfs_v5} shows how this bitmap tracks visited vertices, where the distance array is accessed and updated only when a vertex is found to be unvisited.

\begin{lstlisting}[aboveskip=0em, belowskip=0em, float=!h,label=lst:bfs_v5,caption=Hybrid parallel BFS kernel using bitmap for visited vertex tracking]
// Use a bitmap to check if a vertex is visited
// In top-down
for (uint32_t j = rptr[src]; j < rptr[src+1]; ++j) {
  vidType dst = col[j];
  if (!visited_bm[dst]) {
    dist[dst] = dist[src] + 1;
    lf[lfsz++] = dst;
    visited_bm[dst] = true;
  }
}
// In bottom-up
#pragma omp for nowait reduction(+: nxt_fsz)
for (vidType v = 0; v < N; v++) { 
  if (!visited_bm[v]) { 
    for (uint32_t j = rptr[v]; j < rptr[v+1]; ++j) {
      vidType neighbor = col[j];
      if (curr_front_bm[neighbor]) {
        dist[v] = depth + 1;
        nxt_fsz++;
        nxt_front_bm[v] = true;
        visited_bm[v] = true;
        break;
      }
    }
  }
}
// Optimize further to avoid random accesses. For example, in the top-down phase
for (uint32_t i = 0; i < cfsz; ++i) {
  vidType src = curr_frontier[i];
  for (uint32_t j = rptr[src]; j < rptr[src+1]; ++j) {
    vidType dst = col[j];
    if (!visited_bm[dst]) {
      lf[lfsz++] = dst;
      visited_bm[dst] = true;
    }
  }
}
sort(lf, lf + lfsz);
for (uint32_t i = 0; i < lfsz; ++i) {
  dist[lf[i]] = depth + 1;
}
\end{lstlisting}

This optimization can be extended further to eliminate random memory accesses entirely. The code block between lines 27-41 in Listing~\ref{lst:bfs_v5} demonstrates this approach. The loop now only accesses the visited bitmap. After the loop completes, the local frontier is sorted and distances are updated in order. When the visited bitmap fits in cache, and the size of the local frontier is small, this optimization leads to better cache utilization and fewer memory accesses.

~\\
\noindent \textbf{\texttt{Other Optimizations:}} Among various optimizations we explored, we highlight three key approaches, though none yielded performance gains in our experiments. (a) In the top-down phase, we implemented periodic flushing of local frontiers to the global frontier,  
similar to an optimization in~\cite{beamer:direction-optimizing}. This approach limits local frontier sizes thereby reducing the memory overhead, 
(b) For the bottom-up phase, we employed SIMD instructions to vectorize the neighbor-checking loop and introduced prefetch requests for the \texttt{col} and \texttt{curr\_front\_bm} arrays to minimize cache misses, (c) In the top-down phase, when the global frontier reaches a sufficient size, we distribute work among threads and allow them to process independently without merging local frontiers, either for several iterations or until the global frontier size decreases below a threshold. This optimization reduces synchronization overhead and eliminates bulk memory copies from local to global frontiers. These and additional optimizations we explored will be made available in our public GitHub repository.

%% file: evaluation.tex
\vspace{-0.1in}
\section{Evaluation}
\label{sec:evaluation}

We evaluate the performance of our hybrid BFS algorithm using the graphs listed in Table~\ref{tab:graphs_used_for_evaluation}. Our test suite comprises both synthetic and real-world graphs. Among the synthetic graphs, we include a uniform random graph with $2^{25}$ vertices and a Kronecker graph with $2^{23}$ vertices. Our experiments were conducted on two different platforms:

\sloppy
\begin{hangparas}{1em}{1}
\textbf{SpeedCode Platform~\cite{speedcode}:} The platform employs a \texttt{c5d.12xlarge} instance with an Intel Xeon Platinum 8275L processor at 3.4 GHz. This system features 24 cores, 48 virtual CPUs (vCPUs), and 96 GB of RAM.
\end{hangparas}

\begin{hangparas}{1em}{1}
\textbf{AMD Server:} This machine features a dual-socket AMD EPYC 7V13 system, with each socket housing 64 cores at 2.5 GHz. The system is equipped with 512 GB of RAM. The code is compiled using GCC version 12.2.0 with OpenMP 4.5 support and -O3 optimization level, with thread-to-core pinning enabled. Our experiments with 32, 64, and 128 threads showed optimal performance at 64 threads, which we present in our results.
\end{hangparas}

\begin{figure}[H]
    \centering
    \includegraphics[scale=0.4]{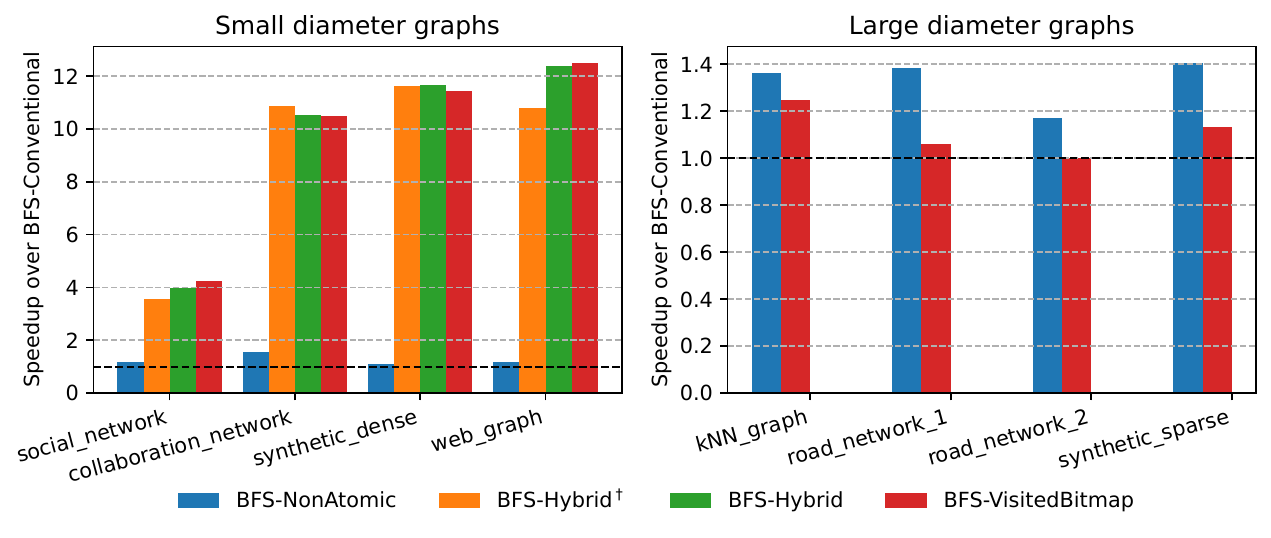}
    \vspace{-5mm}
    \caption{Performance comparison of BFS optimizations on the SpeedCode platform.}
    \label{fig:speedcode_results}
\end{figure}

\begin{figure}[H]
    \centering
    \includegraphics[scale=0.4]{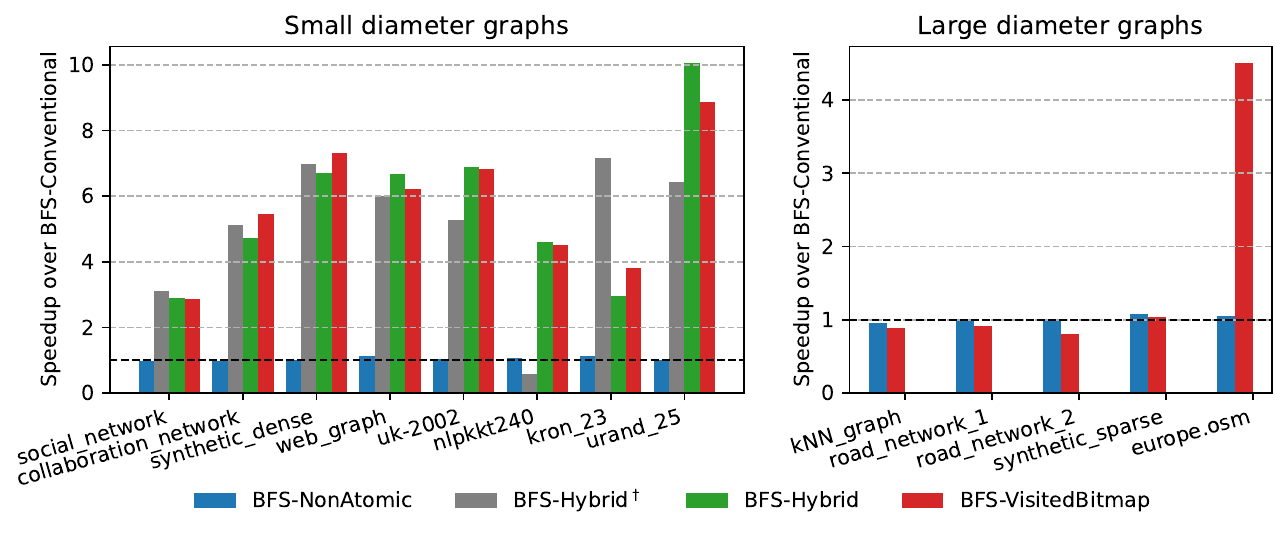}
    \vspace{-5mm}
    \caption{Performance comparison of BFS optimizations on the AMD platform.}
    \label{fig:amd_results}
\end{figure}

We classify graphs into two categories: small-diameter graphs processed using BFS-Hybrid (which switches between top-down and bottom-up phases), and large-diameter graphs processed using only top-down traversal (Figures~\ref{fig:speedcode_results} and~\ref{fig:amd_results}). To reduce synchronization overhead, we implement separate code paths for each category. To efficiently identify graph categories at runtime, we use average-degree as a proxy for diameter, classifying graphs with average-degree less than seven as large-diameter graphs. While this heuristic is generally effective, there are exceptions - for example, \textit{nlpkkt240} has an average-degree of 26 but a diameter of 241 (Table~\ref{tab:graphs_used_for_evaluation}). However, such misclassification does not affect the performance significantly since the BFS-Hybrid algorithm switches between top-down and bottom-up phases as needed.

In our performance evaluation, all plots use \texttt{BFS-Conventional} as the baseline. \texttt{BFS-Conventional} achieves a speedup of $22\times$ and $15.9\times$ when compared to serial BFS on the SpeedCode platform and the AMD system, respectively.

For small-diameter graphs, our optimizations show significant speedups over the conventional BFS implementation across both platforms. On the SpeedCode platform, the \texttt{BFS-Hybrid} and \texttt{BFS-VisitedBitmap} variants achieve speedups of $3-8\times$, while on the AMD system, these speedups increase to $3-10\times$. The performance difference between \texttt{BFS-Hybrid} and its variant shows no clear winner across different graphs. However, we selected \texttt{BFS-Hybrid} as the default choice since it demonstrates better performance on the majority of graphs on the SpeedCode platform.


For large-diameter graphs, our analysis reveals interesting platform-specific characteristics. On the SpeedCode platform, \texttt{BFS-NonAtomic} maintains consistent performance above baseline. However, on the AMD system, \texttt{BFS-NonAtomic} achieves no improvement. This suggests that the trade-off between atomics and duplication of work is platform-dependent. As discussed earlier in our optimization analysis, the overhead of duplicate work may outweigh the benefits of avoiding atomic operations.

The impact of visited bitmap optimization also varies significantly between platforms. While \texttt{BFS-VisitedBitmap} consistently improves performance on small-diameter graphs across both platforms, its behavior on large-diameter graphs differs. On the AMD system, it shows more pronounced performance degradation even to the extent of falling behind the baseline for some graphs. In large-diameter graphs, the neighbors are less in number, and hence the memory overhead of the bitmap might not offset the performance gains of reducing the random memory accesses. \texttt{europe.osm} is a special case where the graph is very sparse, and hence might generate a large number of random memory accesses, leading to a performance improvement with the visited bitmap optimization.

\begin{table}[h]
    \centering
    \begin{tabular}{|l|c|c|c|c|}
        \hline
        Graph & $|V|$ & $|E|$ & Diameter & Degree \\
        \hline
        Collab. Network & 1.1M & 113M   & 9    & 104.29 \\
        Social Network      & 4.9M   & 85.8M  & 14   & 17.5  \\
        Synthetic Dense     & 10M    & 1B     & 6    & 99.01 \\
        Web Graph           & 6.6M   & 300M   & 9    & 44.87 \\
        kNN           & 24.9M  & 158M   & 10945 & 6.27 \\
        Road Network 1      & 22.1M  & 30.0M  & 7479 & 2.66  \\
        Road Network 2      & 87.0M  & 112.9M & 3817 & 2.51  \\
        Synthetic Sparse    & 10M    & 40M    & 5501 & 3.96  \\
        kron\_23               & 8.39M  & 129.34M  & 5   & 15   \\
        urand\_25               & 33.55M & 536.87M  & 6   & 15   \\
        europe.osm    & 50.91M & 57.20M   & 100 & 1    \\
        nlpkkt240     & 27.99M & 746.48M  & 241 & 26   \\
        uk-2002       & 18.52M & 523.57M  & 27  & 28   \\
        \hline
    \end{tabular}
    \caption{Graphs Used for Evaluation}
    \label{tab:graphs_used_for_evaluation}
\end{table}

%% file: conclusion.tex
\section{Conclusion}
We presented a comprehensive study of optimizations for parallel BFS on multicore systems. Our results show that the effectiveness of optimizations is influenced by both graph characteristics and hardware architecture. The bitmap-based vertex tracking optimization demonstrates the importance of memory access patterns, showing consistent improvements for small-diameter graphs but potential degradation for large-diameter graphs. Our simple heuristic of using average degree to classify graphs and select appropriate optimizations proves effective in practice. While parallel BFS is well-studied, our results identify specific opportunities for improvement, particularly in memory access patterns and architecture-specific optimizations. There is potential to explore optimizations targeting emerging hardware features like larger cache hierarchies and heterogeneous computing platforms, as well as more sophisticated graph heuristics, in future work.